\documentclass[submitting]{nst}

\usepackage{subfigure,dcolumn}
\usepackage[T2A,T1]{fontenc}
\usepackage[russian,english]{babel}

\usepackage{listings}
\begin{document}

\title{Shape polarization and coexistence of high-$K$ three-quasiparticle states in odd-mass $N=106$ isotones}\thanks{Supported by the National Natural Science Foundation of China (No.12275369)}

\author{Runyan Dong}
\affiliation{School of Physics and Astronomy, Sun Yat-sen University, Zhuhai 519082, China}
\affiliation{Guangdong Provincial Key Laboratory of Quantum Metrology and Sensing, Sun Yat-sen University, Zhuhai 519082, China}
\author{Changfeng Jiao}
\email[Corresponding author, ]{jiaochf@mail.sysu.edu.cn}
\affiliation{School of Physics and Astronomy, Sun Yat-sen University, Zhuhai 519082, China}
\affiliation{Guangdong Provincial Key Laboratory of Quantum Metrology and Sensing, Sun Yat-sen University, Zhuhai 519082, China}

\begin{abstract}
Three-quasiparticle $K$-isomeric states in odd-mass $N=106$ isotones within the $A\sim 180$ mass region are systematically investigated using configuration-constrained potential energy surface calculations. The calculations successfully reproduce the excitation energies and deformations of known high-$K$ isomers in the nuclei from $^{175}$Tm to $^{181}$Re. For the nuclei closer to the $Z=82$ shell closure ($^{183}$Ir, $^{185}$Au, and $^{187}$Tl), predictions for the configurations of observed and yet-to-be-observed isomers are provided. The results reveal strong shape polarization, where the three-quasiparticle states are driven to larger deformations compared to the often shape-soft or spherical ground states. A particularly rich spectrum of shape coexistence is predicted in $^{187}$Tl, where several high-$K$ three-quasiparticle configurations with distinct prolate, oblate, and triaxial shapes are found to coexist at similar excitation energies. Notably, the oblate-deformed $K^{\pi}=29/2^+$ configuration at $E_x = 1839$ keV is proposed to be responsible for a long-lived isomer. This study provides a comprehensive picture of shape evolution and coexistence in high-$K$ multi-quasiparticle states, offering valuable insights for future experimental research.
\end{abstract}

\keywords{Shape polarization, shape coexistence, high-$K$ isomeric state, configuration-constrained potential energy surface.}

\maketitle
\nolinenumbers

\section{Introduction}

Neutron-deficient isotones around $N=106$ mid-shell are characterized by the existence of an abundance of low-lying, high seniority isomeric states. In this region, the orbitals with large $\Omega$, namely the projection of individual angular momentum onto the intrinsic symmetric axis, approach the neutron Fermi surface at moderate quadrupole deformations. It facilitates the formation of broken-pair states with high $K$ values (where $K =\sum_i\Omega_i$) near the yrast line. According to the  selection rules for electromagnetic transitions, the transitions of multipolarity $\lambda$ would be significantly hindered if $\Delta K > \lambda$. The so-called $K$-hindrance can lead to relatively long half-lives (on the order of nanoseconds or longer)~\cite{walker1999energy,walker2020100,walker2024k}, leading to the formation of ``$K$ isomers''. One of the most well-known examples of high-$K$ isomers is found in $^{178}\mathrm{Hf}$, in which two 2-quasiparticle (qp) $K^\pi = 8^-$ isomers and a long-lived four-quasiparticle $K^\pi = 16^+$ isomer with a half-life of 31 years are observed~\cite{van1980k,mullins1997rotational}. Since the occurrence of $K$ isomers is a combined effect of the unpaired nucleons occupied high-$\Omega$ states and the nuclear deformation, the study of $K$ isomeric states is therefore pivotal for understanding the interplay between the shell structure of ``individual'' nucleons and the collective behavior of a strongly correlated nucleus~\cite{walker1999energy}. 

Another common feature associated with the $A \sim$ 180 mass region is the shape transition. In this region, the ground-state (g.s.) shape can change from a well-deformed prolate ellipsoid with $\beta_2\geqslant0.28$ for $^{176}$Yb to a very soft spheroid for $^{188}$Pb~\cite{xu1999shape,chai2018shape}. Moreover, the soft shape gives rise to the novel shape coexistence phenomenon, which is characterized by the emergence of low-lying states with different intrinsic shapes in one atomic nucleus. In general, it originates from the combining effect of approaching the $Z=82$ spherical shell closures and the deformed shell gaps around the neutron mid-shell at $N = 104-106$ due to quadrupole-quadrupole correlations. 
It has drawn considerable interest ~\cite{heyde1983coexistence, heyde2011shape,garrett2022experimental,guan2025shape}.
The most well-known example is the differently shaped $0^+$ triplet observed in $^{186}$Pb, which corresponds to the coexistence of the prolate, oblate, and spherical configurations~\cite{andreyev2000triplet,ojala2022reassigning}. Coexisting $0^+$ states in even-even Pt, Hg, Pb, and Po isotopes around the neutron midshell have been extensively studied~\cite{heyde2011shape,bengtssonZPA1989,nazarewiczPLB1993,chasmanPLB2001,niksicPRC2002,dracoulis2003isomer,coz2000evidence,egido2004unveiling,oros1999shape,wu2015global}. 

In addition to the shape change resulting from collective correlations such as quadrupole-quadrupole interactions, unpaired nucleons are found to strongly polarize the nuclear shape~\cite{NAZAREWICZ199061}. Since $K$ isomeric states are coupled by high-$\Omega$ unpaired nucleons, shape polarization possibly yields considerable differences in shape between high-$K$ states and ground states, leading to novel structures that involve both $K$ isomerism and shape isomerism. For example, the two-quasineutron $K^\pi=8^- (\nu\left\{7/2^- [514] \otimes 9/2^+ [624]\right\})$ isomeric states which are systematically observed in the even-even $N$=106 isotones between $^{174}\mathrm{Er}$ and $^{188}\mathrm{Pb}$  (see~\cite{walker1994k,dracoulis2009lifetime} and references therein). Previous theoretical investigation has shown that the the g.s. are oblate deformed with $|\beta_2|\approx0.13$ for $^{186}$Hg and spherical for $^{188}$Pb, whereas the $K=8^-$ isomeric states are polarized to prolate deformed with $|\beta_2|\approx0.25$. The $K=8^-$ isomers with shapes different from those of the g.s. have later been confirmed by measuring rotational bands built on them~\cite{PhysRevC.60.014303,PhysRevC.67.051301}. Furthermore, it is found that for shape-soft nuclei, the shape changes, particularly in the triaxial deformations, can be important for understanding the observed behaviors of isomeric states, such as decay properties~\cite{xu1999shape}. 

While the shape evolution and coexistence of high-$K$ states in even-even nuclei around neutron mid-shell and $A\sim180$ have been extensively studied, the structural properties such as the shape changing effects of 3-qp high-$K$ states in their odd-proton neighbors are lacking systematic investigations. In odd-$A$ nuclei, although the unpaired nucleon introduces additional complexity, it also serves as a sensitive probe of the underlying shell structure. The shape polarization effect induced by the single nucleon can be either parallel to or opposed to that of the high-$K$ 2-qp configuration, thereby amplifying or diminishing the shape difference between the 3-qp states and the g.s.. Recently, the 3-qp high-$K$ isomers, originated from the coupling between the odd proton and the aforementioned $K^{\pi} = 8^{-}$ configuration in even-mass cores, have been observed in odd-mass $N = 106$ isotones from $^{175}$Tm to $^{187}$Tl (except the $^{183}$Ir)~\cite{hughes2012high,dracoulis2004identification,barneoud1982multi,pearson2000multi,Byrne2000epja,guo2024spectroscopy,xiangyu2025observation}. In addition, a substantial amount of experimental data also suggest that the low-lying 1-qp states of neutron-deficient odd-mass Au and Tl isotopes exhibit shape coexistence~\cite{heyde1983coexistence,heyde2011shape,garrett2022experimental,venhart2017new,sedlak2020nuclear}. The extent to which the observed 3-qp isomers can be considered to involve shape isomerism, in addition to $K$ isomerism, remains unclear. It thus has greatly stimulated our interest in pursuing theoretical studies on shape polarization and coexistence in the 3-qp high-$K$ states within this mass region.

In this work, we investigate the 3-qp $K$-isomeric states of odd-$A$ nuclei in the $N$ = 106 isotonic chain using the configuration-constrained potential energy surface (PES) method~\cite{xu1998multi}. This method includes the axially asymmetric $\gamma$-degree of freedom. Furthermore, in this method, we do not introduce an adjustable parameter of the pairing strength, and the deformation is determined self-consistently by minimizing the corresponding PES. At prolate deformations, we mainly focus on the study of the high-$K$ 3-qp configurations composed of the coupling of the unpaired proton and the $K^\pi = 8^-$ 2-qp configuration that are observed systematically in even-even $N=106$ nuclei. We predicted the possible configurations of the isomers in $^{185}$Au and $^{187}$Tl, with particular attention to the shape-polarization effect from multi-qp excitations. Furthermore, we explored the high-$K$ states with distinct shapes (oblate, prolate, and triaxial) that coexist at comparable excitation energies in $^{187}$Tl, and analyzed the impact of different quasiparticle configurations on shape evolution in detail.

\section{The model}\label{sec:model}

We employ the configuration-constrained PES approach~\cite{xu1998multi}, based on the macroscopic–microscopic model. The macroscopic energy contribution was computed using the standard liquid-drop model~\cite{myers1966nuclear} with parameters taken from Ref.~\cite{myers1969average}. The microscopic correction includes the Strutinsky shell correction~\cite{strutinsky1967shell} and the pairing correction. The single-particle levels required for the microscopic energy calculations were obtained from a non-axial deformed Woods-Saxon potential~\cite{nazarewicz1985microscopic} using the ``universal'' parameter set~\cite{dudek1981woods}. The so-called universal parameter set (listed in Table~\ref{universalWS}) was optimized by simultaneously fitting the single-particle energies in $^{208}$Pb, particularly those corrected for nucleon-nucleus interactions beyond the independent particle model, as well as the high-spin yrast spectra of $^{212}$Rn and $^{204}$Pb~\cite{dudek1981woods}. These parameters have been further tested for light nuclei~\cite{nazarewicz1985microscopic}, and for heavier nuclei with $63\leqslant Z\leqslant 81$~\cite{NazarewiczProc1987}, showing
satisfactory performance in describing not only
the single (quasi)particle level sequences but also
the nuclear equilibrium deformations. Since these parameters were optimized for the lead region, they have been validated as the best-fit Woods-Saxon potential parameters for this mass region.

\begin{table}[!htbp]
\renewcommand\arraystretch{1.5}
\centering
\caption{\label{universalWS}The universal parameters of the Woods-Saxon potential}
\resizebox{\columnwidth}{!}{
\begin{ruledtabular}
\begin{tabular}{*{11}{c}}
& Particles  &  $V_0$  & $\kappa$  & $a$ & $r_n$  &  $\lambda_n$  & $r_n^{so}$ \\
\hline
& Neutron  & 49.6 & 0.86 & 0.70 & 1.347 & 35.0 & 1.31  \\
& Proton  & 49.6 & 0.86 & 0.70 & 1.275 & 36.0  & 1.30 \\
\end{tabular}
\end{ruledtabular}
}
\end{table}

To avoid the collapse of pairing correlations in multi-quasiparticle states, we used the Lipkin-Nogami (LN) version of the Bardeen-Cooper-Schrieffer (BCS) method~\cite{pradhan1973study} as an approximate particle-number projection scheme, incorporating monopole pairing. The pairing strength $G$ was initially determined via the average-gap method~\cite{moller1992nuclear,moller1993nuclear}. Although it is often further adjusted to reproduce the odd-even mass difference using a five-point formula, we note that irregularities may arise near magic numbers (e.g., Au and Tl isotopes)~\cite{moller1992nuclear}. This is partly due to the limitation of the BCS approach and the artefacts from the LN correction. It is also because the
odd-even mass difference equations used to extract experimental pairing gaps are derived under the assumption that there are no non-smooth contributions to
the masses apart from pairing effects, while this assumption is often not fulfilled at magic numbers~\cite{moller1992nuclear}. Therefore, following the recommendation of Ref.~\cite{moller1992nuclear}, closed-shell nuclei were excluded from the pairing strength calibration. For consistency, we adopted the standard pairing strength across all isotopes under investigation.

In the PES calculations, a deformation mesh in $\left(\beta_{2}, \gamma\right)$ is used, with the hexadecapole deformation $\beta_{4}$ variation at each mesh point. 
For broken-pair configurations, the microscopic energy incorporates contributions from unpaired nucleons occupying specific single-particle orbitals (see Ref.~\cite{xu1998multi} for details). These orbitals are continuously tracked and adiabatically blocked throughout the $(\beta_2,\gamma)$ deformation plane. For axially deformed nuclei, single-particle orbitals can be specified by their individual spin projection $\Omega$. For axially asymmetric shapes, however, $\Omega$ is no longer conserved. In addition, several single-particle orbits with approximately the same spin projection $|\Omega|$ may become energetically close at given $(\beta_2,\gamma)$ deformations. To reliably track orbits, we computed and compared not only the average spin projection $|\Omega|$ but also the expectation values of the Nilsson numbers $\langle N \rangle$, $\langle n_z \rangle$, and $\langle \Lambda \rangle$ between two adjacent deformation mesh points. We have verified that, although the Nilsson numbers $N$, $n_z$, $\Lambda$, and $|\Omega|$ are not strictly conserved, their expectation values exhibit slow variation, allowing for a reliable configuration assignment. Therefore, each configuration is identified by computing the average Nilsson quantum numbers of the blocked orbitals. 
The total energy of a multi-qp state with unpaired nucleons can be decomposed into the deformation energy and the configuration energy, where the latter originates from qp excitations due to pair breaking and excitations of particles that define the specific configuration.

Quasiparticle excitations, particularly in deformation-soft nuclei, can induce significant shape polarization, resulting in an equilibrium deformation for the multi-qp state that differs from that of the ground state. The configuration-constrained PES method effectively accounts for this polarization caused by the unpaired nucleon and offers a self-consistent description of both the deformation and excitation energy of multi-qp states~\cite{xu1999shape,shi2010multi}. The excitation energy is computed as the energy difference between the PES minimum of the excited configuration and that of the ground-state configuration, enabling direct comparison with experimental values.

\section{Calculations and discussions}
\subsection{Systematics of 3-qp states involved $\nu \left\{9/2^{+}[624] \otimes 7/2^{-}[514]\right\}$}\label{systematics}

\begin{table*}[!htbp]
\renewcommand\arraystretch{1.5} 
\setlength{\arrayrulewidth}{0.3mm} 
\centering
\caption{\label{table1}Calculated deformations $\left(\beta_{2}, \beta_{4}, \gamma\right)$ and excitation energies $E_{\mathrm{cal}}$ for odd-$A$ nuclei in the $N$ = 106 isotopic chain. The experimental energies $E_{\mathrm{exp}} $ can be found in Refs.~\cite{kondev2015configurations,guo2024spectroscopy} and references therein.}
\begin{ruledtabular}
\begin{tabular}{cllrrcrr}
   Nuclei & $K^{\pi}$ & Configuration & $E_{\mathrm{exp}} (\mathrm{keV})$ & $E_{\mathrm{cal}} (\mathrm{keV})$ & $\beta_{2}$   & $\beta_{4}$ & $\gamma$  \\
    \hline
    ${ }^{175} \mathrm{Tm}$  &  $\frac{1}{2}^{+}(\mathrm{g.s.})$ &  $\pi \frac{1}{2}^{+}[411]$ & 0 & 0 & 0.279  &  $-$0.042  & 0.04  \\
    &  $\frac{7}{2}^{-}$ &  $\pi \frac{7}{2}^{-}[523]$ & 439 & 28 & 0.276 & $-$0.041 & 0.08  \\
    &  $\frac{17}{2}^{-}$  &  $\pi \frac{1}{2}^{+}[411] \otimes \nu \left\{ \frac{9}{2}^{+}[624] \otimes \frac{7}{2}^{-}[514]\right\}$  &  1004.8 &  1176 &  0.281  & $-$0.042 & 0.06  \\
    &  $\frac{23}{2}^{+}$  &  $\pi \frac{7}{2}^{-}[523] \otimes \nu  \left\{\frac{9}{2}^{+}[624] \otimes \frac{7}{2}^{-}[514]\right\}$  &  1517.7 &  1343 &  0.279  & $-$0.043 & 0.06 \\[8pt]
    ${ }^{177} \mathrm{Lu}$  & $\frac{7}{2}^{+}(\mathrm{g.s.})$ &  $\pi \frac{7}{2}^{+}[404]$ & 0 & 259 &  0.257  & $-$0.043 & $-0.05$  \\ 
    & $\frac{9}{2}^{-}$ &  $\pi \frac{9}{2}^{-}[514]$ & 150.4 & 0 & 0.270  &  $-$0.055  & 0.06 \\
    & $\frac{23}{2}^{-}$  &  $\pi \frac{7}{2}^{+}[404] \otimes \nu \left\{ \frac{9}{2}^{+}[624] \otimes \frac{7}{2}^{-}[514]\right\}$  &  970.2 &  1487 &  0.261  & $-$0.043 & 0.06 \\
    &  $\frac{25}{2}^{+}$  &  $\pi \frac{9}{2}^{-}[514] \otimes \nu \left\{\frac{9}{2}^{+}[624] \otimes \frac{7}{2}^{-}[514]\right\}$  &  1324.4 & 1197 &  0.273 & $-$0.054 & 0.05 \\[8pt]
   
     ${ }^{179} \mathrm{Ta}$ &  $\frac{7}{2}^{+}(\mathrm{g.s.})$ &  $\pi \frac{7}{2}^{+}[404]$ & 0 & 0 &  0.248 & $-$0.049  & 0.04  \\
     &  $\frac{9}{2}^{-}$ &  $\pi \frac{9}{2}^{-}[514]$ & 30.7 & 16 & 0.245 & $-$0.047  & 0.00 \\
      &  $\frac{21}{2}^{-}$ & $\pi \left\{\frac{7}{2}^{+}[404] \otimes \frac{5}{2}^{+}[402] \otimes \frac{9}{2}^{-}[514] \right\}$ & 1252.6 & 1313  & 0.241 &  $-$0.049  & $-0.09$ \\
      &  $\frac{23}{2}^{-}$ & $\pi \frac{7}{2}^{+}[404] \otimes \nu \left\{\frac{9}{2}^{+}[624] \otimes \frac{7}{2}^{-}[514]\right\}$ & 1328  &  1260  & 0.254 & $-$0.048 &  0.17  \\
      &  $\frac{25}{2}^{+}$ & $\pi \frac{9}{2}^{-}[514] \otimes \nu \left\{\frac{9}{2}^{+}[624] \otimes \frac{7}{2}^{-}[514]\right\}$ & 1317 & 1279 & 0.254 &  $-$0.046  &  0.13 \\[8pt]
      ${ }^{181} \mathrm{Re}$  &  $\frac{5}{2}^{+}(\mathrm{g.s.})$ &  $\pi \frac{5}{2}^{+}[402]$ & 0 & 0 & 0.219  &  $-$0.045  & $-0.18$  \\
      &  $\frac{9}{2}^{-}$ &  $\pi \frac{9}{2}^{-}[514]$ & 262.9 & 201 &  0.216 & $-$0.040  &  $-0.63$ \\
      &  $\frac{21}{2}^{-}$  &  $\pi \frac{5}{2}^{+}[402] \otimes \nu \left\{\frac{9}{2}^{+}[624] \otimes \frac{7}{2}^{-}[514]\right\}$  &  1656.4  &  1657  &  0.230  &  $-$0.043  & $-0.36$ \\ 
       &  $\frac{25}{2}^{+}$ & $\pi \frac{9}{2}^{-}[514] \otimes \nu \left\{\frac{9}{2}^{+}[624] \otimes \frac{7}{2}^{-}[514]\right\}$ & 1880.6 & 1864 & 0.229  & $-$0.037 &  $-0.48$ \\[8pt]
       ${ }^{183} \mathrm{Ir}$  &  $\frac{1}{2}^{-}(\mathrm{g.s.})$ &  $\pi \frac{1}{2}^{-}[541]$ &  & 0 & 0.224  &  $-$0.030  & 0.02  \\
       &  $\frac{9}{2}^{-}$ &  $\pi \frac{9}{2}^{-}[514]$ & 645.3 & 492 & 0.222  &  $-$0.023  & $-2.10$  \\
      &  $\frac{17}{2}^{+}$  &  $\pi \frac{1}{2}^{-}[541] \otimes \nu \left\{\frac{9}{2}^{+}[624] \otimes \frac{7}{2}^{-}[514]\right\}$  &    &  1552  &  0.233  &  $-$0.030  & 0.03 \\ 
       &  $\frac{25}{2}^{+}$ & $\pi \frac{9}{2}^{-}[514] \otimes \nu \left\{\frac{9}{2}^{+}[624] \otimes \frac{7}{2}^{-}[514]\right\}$ &  & 2011 & 0.245  & $-$0.025 &  0.05 \\[8pt]
       ${ }^{185} \mathrm{Au}$ &  $\frac{1}{2}^{-}(\mathrm{g.s.})$ &  $\pi \frac{1}{2}^{-}[541]$ &  & 0 & 0.151  &  $-$0.025 & 23.53  \\
        & $\frac{1}{2}^{+}$ &  $\pi \frac{1}{2}^{+}[411]$ & 24 & 169  &  0.183 & $-0.001$  & $-59.08$ \\
       &  $\frac{3}{2}^{-}$ &  $\pi \frac{3}{2}^{-}[532]$ &  & 523 &  0.235 &  $-$0.029  & 11.70 \\
       &  $\frac{9}{2}^{-}$ &  $\pi \frac{9}{2}^{-}[505]$ & 8.9 & 269 &  0.198 &  $-$0.022  & $-24.84$ \\
       &  $\frac{17}{2}^{+}$  &  $\pi \frac{1}{2}^{-}[541] \otimes \nu \left\{\frac{9}{2}^{+}[624] \otimes \frac{7}{2}^{-}[514]\right\}$  &   &  1967  & 0.230  &  $-$0.033  & 1.04  \\
       &  $\frac{19}{2}^{+}$  &  $\pi \frac{3}{2}^{-}[532] \otimes \nu \left\{\frac{9}{2}^{+}[624] \otimes \frac{7}{2}^{-}[514]\right\}$ & &  1981  & 0.246  & $-$0.031  & 1.14 \\[8pt]
      ${ }^{187} \mathrm{Tl}$  & $\frac{1}{2}^{+}(\mathrm{g.s.})$ &  $\pi \frac{1}{2}^{+}[400]$ & 0 & 0 & 0.024  & 0.001  & $-2.27$  \\
      & $\frac{11}{2}^{-}$ &  $\pi \frac{11}{2}^{-}[505]$ & 952 & 888 &  0.214 &  $-$0.014 & 18.32 \\
      &  $\frac{27}{2}^{+}$  &  $\pi \frac{11}{2}^{-}[505] \otimes \nu \left\{\frac{9}{2}^{+}[624] \otimes \frac{7}{2}^{-}[514]\right\}$  &  &  2312  & 0.225 & $-$0.016  & $-12.13$ \\
\end{tabular}
\end{ruledtabular}
\end{table*}

For nuclei in the $A \sim 180$ region, an abundance of high-$K$ isomeric states has been discovered~\cite{kondev2015configurations}. Among them, the two-quasineutron $K^\pi = 8^- (\nu\left\{7/2^- [514] \otimes 9/2^+ [624]\right\})$ isomeric states exist systematically in even-even $N$ = 106 isotones, which have been investigated by means of the configuration-constrained PES calculation in Ref.~\cite{xu1999shape}. The calculated excitation energies agree well with the experimental data, and strong shape polarizations have been found when approaching the $Z=82$ shell closure. For odd-mass $N = 106$ isotones in this mass region, most of the observed 3-qp $K$ isomers consistently involve a two-quasineutron configuration coupled to $K^\pi = 8^- $ states that are identified in the aforementioned even-even nuclei. For example, in nuclei such as $^{175}$Tm~\cite{lovhoiden1979single,hughes2012high}, $^{177}$Lu~\cite{dracoulis2004identification, kondev2012m}, $^{179}$Ta~\cite{barneoud1982multi, kondev1997multi, kondev2004k}, and $^{181}$Re~\cite{pearson2000multi, pfutzner2002angular}, 3-qp $K$ isomers have been assigned as the two-quasineutron $K^\pi = 8^- $ configuration coupled to the energetically lowest one-quasiproton configuration. Furthermore, in $^{181}$Re, $^{179}$Ta, and $^{177}$Lu, the meta-stable $K^{\pi}=9/2^-$ states are found, assigned to the $\pi9/2^- [514]$ configuration, leading to the presence of 3-qp states associated with the coupling of a $K^\pi = 8^- $ two-quasineutron configuration with $\pi9/2^- [514]$~\cite{dracoulis2004identification,barneoud1982multi,pearson2000multi}. In $^{175}$Tm, a $K$ isomer which may involve coupling of the two-quasineutron $K^\pi = 8^- $ configuration with the $\pi 7/2^- [523]$ has been found~\cite{hughes2012high}. The half-lives of these isomers range from a few microseconds to several days. To what extent these high-$K$ states are associated with shape polarization and shape isomerism is still an open question.
 
We have performed the configuration-constrained PES calculations on the 1-quasiproton and 3-qp states in $N$ = 106 odd-mass isotones. Table \ref{table1} presents the calculated deformations and energies of the g.s., the possible high-$\Omega$ 1-quasiproton and low-lying high-$K$ 3-qp states, compared with the available experimental data. 
Our calculations reproduce the experimentally assigned spin-parity of the g.s. of these nuclei, except for $^{177}$Lu, in which the calculated lowest 1-quasiproton configuration is the $\pi9/2^{-}[514]$ rather than the experimentally assigned $\pi7/2^{+}[404]$~\cite{georgepjalu177}. However, the calculated $\pi7/2^{+}[404]$ configuration lies only 259 keV above the $\pi9/2^{-}[514]$ state. Given the strong dependence of 1-qp state energies on the ordering and spacing of single-particle levels, the deviation in their relative positions falls within an acceptable range. 

One may ask the sensitivity of these calculated results to the pairing strengths or the Woods–Saxon parameters. Note that previous work has shown that adjustment of pairing strength mainly influences the quasiparticle energy gaps and only slightly affects the deformations~\cite{xu1998multi}, while the ordering of single-particle levels at a certain deformation is primarily determined by the choice of the Woods–Saxon potential parameters. To examine the robustness of this result, we therefore have tested with several known parameter sets, including those of Blomquist and Wahlborn~\cite{Blomquist1960}, the parameters of Chepurnov~\cite{Chepurnov1967}, the parameters given by Rost~\cite{RostPLB1968}, and the ``new'' parameters~\cite{DudekJPG1979}. We found that the deviation in the relative positions of the $\pi9/2^{-}[514]$ and $\pi7/2^{+}[404]$ configurations of $^{177}$Lu is preserved with different parameter sets. To accurately reproduce the ordering of the low-lying 1-qp states of $^{177}$Lu, an improvement of the Woods-Saxon potential is needed.

Experimentally, the spin and parity of the g.s. of $^{183}$Ir~\cite{ladenbauer1975some} and $^{185}$Au~\cite{desthuilliers1979structure,BERG1983445,larabee1986shape} have been assigned to be the $5/2^{-}$ states built on the $\pi1/2^-[541]$ configuration, while a strong mixing between the $\pi1/2^-[541]$ and $\pi3/2^-[532]$ configurations attributed to the Coriolis interactions has been proposed for the $5/2^-$ g.s. of $^{185}$Au ~\cite{WALLMEROTH1989224}. Our calculations show that the $\pi1/2^-[541]$ configuration has the lowest energy, while the $\pi3/2^-[532]$ state is about 500 keV higher. The present PES calculations show that these two low-lying 1-quasiproton states both have considerable triaxial deformations, which would reinforce substantial configuration mixing. However, configuration mixing calculations are beyond the scope of the present work.

The present configuration-constrained PES calculations also reasonably reproduce the high-$\Omega$ 1-qp isomeric states observed in odd-mass $N=106$ isotones except for the aforementioned deviation in $^{177}$Lu. Notably, the configuration-constrained PES calculations show that the $K^{\pi}=9/2^-$ isomer of $^{185}$Au and the $K^{\pi}=11/2^-$ state of $^{187}$Tl have moderate triaxial deformations with $\beta_2\sim 0.2$ which are remarkably polarized when compared with their g.s.. It indicates the appearance of single-proton-induced shape polarization in shape-soft odd-proton nuclei when approaching the $Z=82$ closed shell. 

Now we turn to the investigation of energetically low-lying 3-qp states in odd-mass $N=106$ isotones. We mainly focus on the 3-qp states that consist of the two-quasineutron $\nu\left\{7/2^- [514] \otimes 9/2^+ [624]\right\}$ configuration coupled to different 1-quasiproton configurations, since the $K^{\pi}=8^-$ isomeric states are systematically identified as the $\nu\left\{7/2^- [514] \otimes 9/2^+ [624]\right\}$ configuration in even-even $N=106$ isotones. As seen in Table~\ref{table1}, the calculated energies of these 3-qp states in $^{175}$Tm, $^{179}$Ta, and $^{181}$Re agree well with the experimental data. For $^{177}$Lu, the calculated $K^{\pi}=23/2^{-}$ $\pi 7/2^{+}[404] \otimes \nu \left\{ 9/2^{+}[624] \otimes 7/2^{-}[514]\right\}$ configuration is overpredicted, while the $K^{\pi}=25/2^{+}$ $\pi 9/2^{-}[514] \otimes \nu \left\{ 9/2^{+}[624] \otimes 7/2^{-}[514]\right\}$ state is slightly underestimated. This can be attributed to the deviation of the $\pi 7/2^{+}[404]$ and $\pi 9/2^{-}[514]$ orbitals that we found in the calculations of the 1-quasiproton states.

\begin{figure}[t]
\hspace*{-0.05\columnwidth}
\includegraphics[width=1\columnwidth]{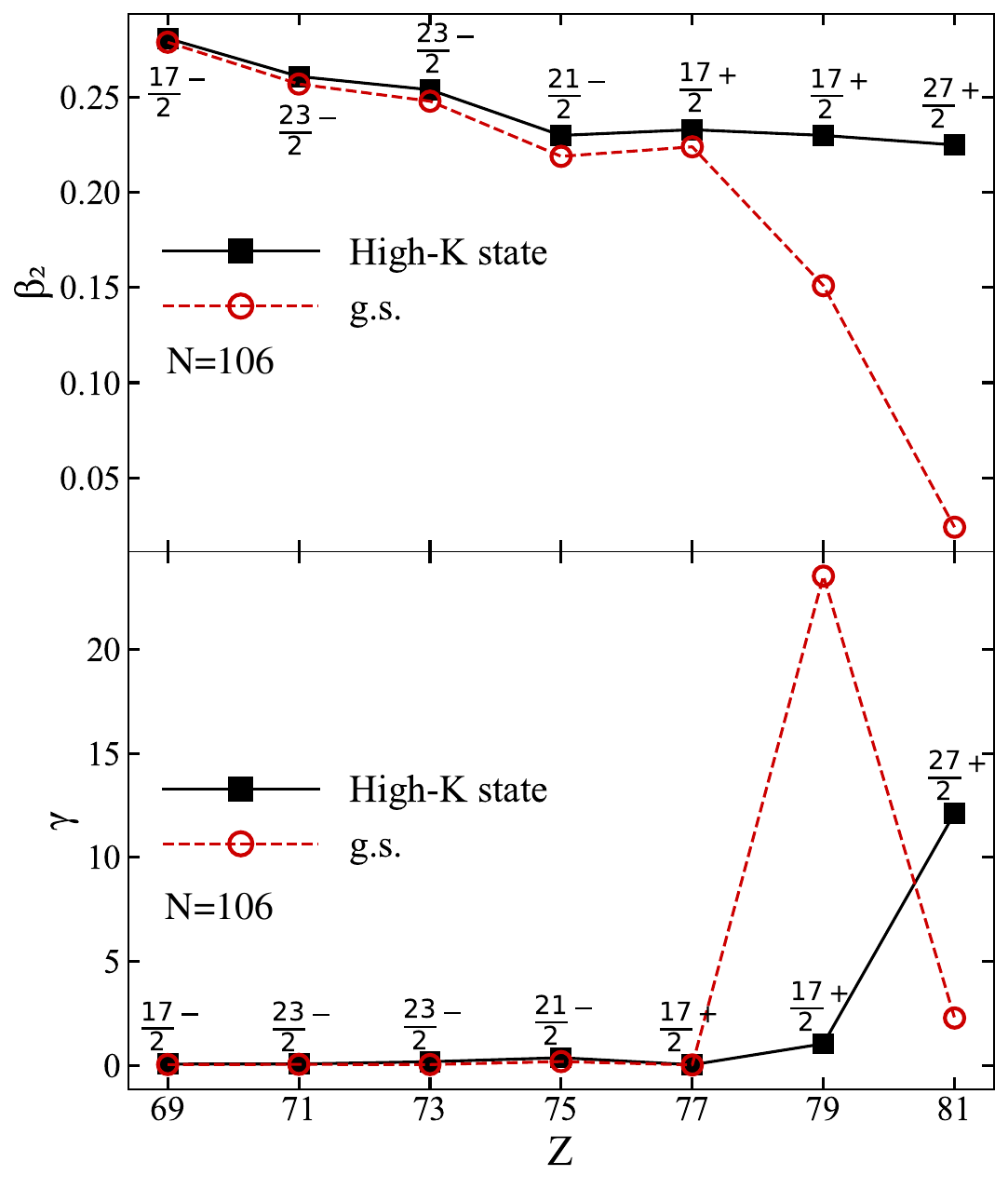}
\caption{Calculated $\beta_2$ and $\gamma$ deformations for the g.s. and 3-qp states of odd-$A$ nuclei in the $N$=106 isotonic chain. For $Z=69{-}79$, the 3-qp states correspond to the coupling of the 2-qp $K^{\pi}=8^{-}$ configuration $\nu7/2^- [514] \otimes 9/2^+ [624]$ with the g.s. configurations of odd-$A$ nuclei, which are $\pi1/2^{+}[411]$, $\pi7/2^{+}[404]$, $\pi7/2^{+}[404]$, $\pi5/2^{+}[402]$, $\pi1/2^{-}[541]$, and $\pi1/2^{-}[541]$, respectively. For $Z=81$, the 3-qp state is $\pi {11/2}^{-}[505] \otimes \nu \{ {9/2}^{+}[624] \otimes {7/2}^{-}[514] \}$. These 3-qp states correspond to observed high-$K$ isomers.}\label{fig1}
\end{figure}

Furthermore, the configuration-constrained PES calculations predict the candidate configurations of the 3-qp isomeric states in the odd-$A$ $N=106$ isotones when moving towards the $Z=82$ shell closure. To date, no experimental evidence has been reported on three-quasiparticle high-$K$ isomers in $^{183}$Ir. We proposed two possible high-$K$ 3-qp states that are composed of two-quasineutron $\nu\left\{7/2^- [514] \otimes 9/2^+ [624]\right\}$ coupled with the proton configurations of $\pi1/2^{-}[541]$ and $9/2^{-}[514]$, respectively. For $^{185}$Au, a new
isomer at an excitation energy of 1504.2(4) keV with a half-life of 630(80) ns was identified in
$\gamma-\gamma$ coincidence analysis recently~\cite{xiangyu2025observation}. The possible spins of this isomer are constrained to a range from 13/2 to 21/2 in comparison with predictions from the Weisskopf estimates~\cite{xiangyu2025observation}. Our calculations suggest two possible 3-qp high-$K$ configurations that are consistent with systematics of 3-qp configurations in lighter odd-mass $N=106$ isotones. They both lie at excitation energies of about 1900 keV, which is a bit overpredicted. However, Ref.~\cite{xiangyu2025observation} argued that the g.s. configuration of $^{185}$Au is more likely $\pi3/2^{-}[532]$, and the calculated energy differences of these two 3-qp states with respect to the $3/2^{-}[532]$ configuration are 1444 and 1458 keV, respectively, which is in great agreement with the observation. For nucleus $^{187}$Tl, two isomers with microsecond lifetimes ($T_{1/2} = 1.11$ $\mu$s and 0.69 $\mu$s) have been
reported~\cite{Byrne2000epja}. Spin-parities $J^{\pi} = 27/2^+, 31/2^-$ are tentatively assigned to the isomer lies at 2584 keV with lifetime $T_{1/2} = 0.69$ $\mu$s based on the deduced total conversion coefficient~\cite{guo2024spectroscopy}. Our calculation presents a $K^{\pi}= 27/2^+$, $\pi 11/2^{-}[505] \otimes \nu \left\{9/2^{+}[624] \otimes 7/2^{-}[514]\right\}$ configuration with an excitation energy of 2312 keV, which is in accord with the prolate high-$K$ configuration suggested in Ref.~\cite{Byrne2000epja}. This implies that the $T_{1/2} = 0.69$ $\mu$s isomer observed in $^{187}$Tl involves the two-quasineutron $\nu \left\{9/2^{+}[624] \otimes 7/2^{-}[514]\right\}$ configuration, which is consistent with the systematics of 3-qp isomers observed in lighter odd-mass $N=106$ isotones. However, further experimental data are required to unambiguously assign
the spin-parity and configuration of these observed isomers. We would discuss the other $T_{1/2} = 1.1$ $\mu$s isomer later in Sect.~\ref{shapecoexistence}. 

In addition, intrinsic shape evolution is crucial for understanding the observed behavior of isomeric states, such as their decay properties. In the previous study~\cite{xu1999shape}, strong shape polarization has been shown in even-even nuclei with $A\sim180$ and $N=106$, especially in nuclei close to the $Z = 82$ shell closure. For systematic comparison, we plot the variation of $\beta_2$ and $\gamma$ deformations of the high-$K$ 3-qp states and the g.s. along with the proton number $Z$ in Fig.~\ref{fig1}. When approaching the shell closure of $Z = 82$, the $\beta_2$ value of the g.s. gradually decreases, indicating that the g.s. shape evolves towards a spheroid, whereas the 3-qp states are polarized to have distinct prolate shapes. The g.s. of $^{185}$Au, for example, has a remarkably $\gamma$-soft shape with a very shallow PES minimum at $\gamma\approx 24^{\circ}$. Considering that the g.s. of $^{185}$Au can be interpreted as the coupling of a low-$\Omega$ $\pi 1h_{9/2}$ proton with the $^{184}$Pt core, the PES of the g.s. of $^{185}$Au exhibits similar $\gamma$ softness compared to $^{184}$Pt. This $\gamma$-soft nature of $^{184}$Pt is mainly because the prolate-to-oblate shape transition going through a transitional $\gamma$-soft shape
occurs in Pt isotopes around $N=110$, which has been suggested by the self-consistent HFB calculation using Gogny-D1S interaction and the interacting boson model~\cite{Garcia-RamosPRC2014}, as well as the five-dimensional collective Hamiltonian (5DCH) based on covariant density-functional theory~\cite{YangPRC2017}. In contrast, the predicted two 3-qp high-$K$ states of $^{185}$Au both have an approximately prolate shape with $\gamma\approx 1^{\circ}$ and about a $60\%$ increase in $\beta_2$ deformation. The nucleus $^{187}$Tl has a spherical g.s. with a proton that singly occupies the $\pi3s_{1/2}$ orbital, while the calculated $K^{\pi}=27/2^+$ $\pi 11/2^{-}[505] \otimes \nu \left\{9/2^{+}[624] \otimes 7/2^{-}[514]\right\}$ state is predicted to have a moderately axially-asymmetric shape with $\beta_2\approx0.23$ and $|\gamma|\approx12^{\circ}$, exhibiting the greatest difference in quadrupole deformation between the 3-qp state and the g.s.. In fact, the calculated high-$K$ 3-qp configurations of $^{187}$Tl present an ensemble of multiple nuclear shapes, which would be analyzed in detail in Sect.~\ref{shapecoexistence}.

\subsection{Shape coexistence in high-$K$ 3-qp states of $^{187}$Tl}\label{shapecoexistence}

The shape-coexisting configurations in this mass region are mainly attributed to the large spherical and deformed shell gaps that simultaneously appear near the proton shell closure at $Z=82$ and the neutron mid-shell at $N \approx 106$. Moreover, the unpaired nucleon that occupies different single-particle orbitals would polarize the shape of the odd-$A$ nucleus in different ways, leading to more profound shape coexistence phenomenon.   
An abundance of experimental data has demonstrated that differently-shaped configurations in these odd-$A$ nuclei are observed not only in the low-lying 1-qp states but also in the higher-seniority isomeric states~\cite{heyde2011shape}.

\begin{table*}[ht]
\renewcommand\arraystretch{1.5} 
\setlength{\arrayrulewidth}{0.3mm} 
\centering
\caption{\label{table2}Calculated deformations $\left(\beta_{2}, \beta_{4}, \gamma\right)$ and excitation energies $E_{\mathrm{cal}}$ for the g.s., the low-lying high-$\Omega$ 1-qp, and high-$K$ 3-qp states of $^{187}$Tl. The experimental data can be found in Refs.~\cite{Byrne2000epja,guo2024spectroscopy} and references therein.}
\begin{ruledtabular}
\begin{tabular}{cllrrcrr}
   & $K^{\pi} $  &  Configuration   & $E_{\mathrm{exp}}(\mathrm{keV})$ &  $E_{\mathrm{cal}}(\mathrm{keV})$  & $\beta_{2}$   & $\beta_{4}$  &  $\gamma$   \\
    \hline
    &  $\frac{1}{2}^{+}(\mathrm{g.s.})$ &  $\pi \frac{1}{2}^{+}[400]$ & 0 & 0 & 0.024  & 0.001  & $-2.27$  \\
    &  $\frac{9}{2}^{-}$ &  $\pi \frac{9}{2}^{-}[505]$ & 335 & 147 & 0.164  &  $-$0.005  & $-59.62$  \\
    &  $\frac{11}{2}^{-}$ &  $\pi \frac{11}{2}^{-}[505]$ & 952 & 888 & 0.214  &  $-$0.014  & 18.32  \\
    &  $\frac{13}{2}^{+}$ &  $\pi \frac{13}{2}^{+}[606]$ & 1061 & 1202 & 0.184  & 0.007  & $-59.95$ \\
    &  $\frac{1}{2}^{+}$ &  $\pi \frac{1}{2}^{+}[660]$ &  & 1216 & 0.267  & $-$0.012  &  $-17.49$ \\
    &  $\frac{25}{2}^{-}$  &  $\pi \frac{9}{2}^{-}[505] \otimes \nu \left\{\frac{9}{2}^{+}[624] \otimes \frac{7}{2}^{+}[633]\right\}$  & & 742 &  0.168 & $-$0.019 & $-59.89$ \\
    &  $\frac{29}{2}^{+}$  &  $\pi \frac{13}{2}^{+}[606] \otimes \nu \left\{\frac{9}{2}^{+}[624] \otimes \frac{7}{2}^{+}[633]\right\}$  & & 1839 & 0.192 & $-$0.018 & $-59.46$ \\
    &  $\frac{27}{2}^{+}$  &  $\pi \frac{11}{2}^{-}[505] \otimes \nu \left\{\frac{9}{2}^{+}[624] \otimes \frac{7}{2}^{-}[514]\right\}$  &  &  2312  & 0.225 & $-$0.016  & $-12.13$\\
    &  $\frac{25}{2}^{+}$  &  $\pi \frac{9}{2}^{-}[505] \otimes \nu \left\{\frac{11}{2}^{+}[615] \otimes \frac{5}{2}^{+}[642]\right\}$  & & 2518 & 0.147 & 0.001 & $-63.46$ \\
    &  $\frac{29}{2}^{-}$  &  $\pi \frac{9}{2}^{-}[505] \otimes \nu \left\{\frac{9}{2}^{+}[624] \otimes \frac{11}{2}^{+}[615]\right\}$  & & 2543 &  0.154 & 0.000 & $-59.83$ \\
    &  $\frac{25}{2}^{+}$  &  $\pi \frac{11}{2}^{-}[505] \otimes \nu \left\{\frac{9}{2}^{+}[624] \otimes \frac{5}{2}^{-}[512]\right\}$  & & 2562 & 0.224 & $-$0.016 & 12.10 \\
    &  $\frac{25}{2}^{+}$  &  $\pi \frac{11}{2}^{-}[505] \otimes \nu \left\{\frac{7}{2}^{+}[633] \otimes \frac{7}{2}^{-}[503]\right\}$  & & 2630 &  0.203 &  0.003  & $-38.51$ \\
    &  $\frac{25}{2}^{-}$  &  $\pi \frac{11}{2}^{-}[505] \otimes \nu \left\{\frac{9}{2}^{-}[505] \otimes \frac{7}{2}^{-}[514]\right\}$  & & 2835 & 0.209 & $-$0.007 & $-20.82$ \\
    &  $\frac{27}{2}^{-}$  &  $\pi \frac{11}{2}^{-}[505] \otimes \nu \left\{\frac{9}{2}^{-}[505] \otimes \frac{7}{2}^{-}[503]\right\}$  & & 2933 & 0.180 & $-$0.004 & $-29.17$ \\
    &  $\frac{27}{2}^{+}$  &  $\pi \frac{11}{2}^{-}[505] \otimes \nu \left\{\frac{9}{2}^{-}[505] \otimes \frac{7}{2}^{+}[633]\right\}$  &  & 2939 & 0.181 & $-$0.007 & $-28.83$ \\
    &  $\frac{25}{2}^{+}$  &  $\pi \frac{11}{2}^{-}[505] \otimes \nu \left\{\frac{7}{2}^{-}[514] \otimes \frac{7}{2}^{+}[633]\right\}$  & & 2962 & 0.214 & $-$0.014 & $-15.55$ \\
    &  $\frac{29}{2}^{+}$  &  $\pi\left\{ \frac{13}{2}^{+}[606] \otimes \frac{9}{2}^{-}[505] \otimes \frac{7}{2}^{-}[514]\right\}$  & & 2983 & 0.235 & 0.004 & $-58.26$ \\
\end{tabular}
\end{ruledtabular}
\end{table*}

For $^{187}$Tl, previous studies~\cite{lane1995shape,lee2012observation,chamoli2005shape,reviol1994prolate} have proposed the coexistence of different nuclear shapes by the analysis of observed low-lying collective structures. As the proton-hole neighbor of $^{188}$Pb, the $I=1/2^+$ g.s. of $^{187}$Tl can be interpreted as the coupling of the $\pi3s_{1/2}$ hole with the spherical $0_1^+$ state of $^{188}$Pb core. The observed $K^{\pi}=9/2^-$ and $K^{\pi}=13/2^+$ isomeric states can be understood as filling the $\pi9/2^-[505]$ and $\pi13/2^+[606]$ intruder orbitals that are lowered along with the increase in oblate deformation, respectively~\cite{barzakh2013changes,barzakh2017changes}, while the rotational band built on the $I^{\pi}=11/2^-$ state is suggested to be the prolate deformed $\pi 11/2^-[505]$ configuration~\cite{Byrne2000epja,guo2024spectroscopy}. The calculated deformations and excitation energies for these 1-quasiproton states are listed in Table~\ref{table2}. The configuration-constrained PES calculations well reproduce the measured excitation energies, and clearly show the coexisting shapes for these 1-qp states. Note that the $\pi 11/2^-[505]$ configuration has been predicted to have a considerable axial-asymmetric shape of $\gamma\approx 18^{\circ}$, which breaks down the $K$- and shape-hindrance and would explain why it decays fast to the oblate $K^{\pi}=9/2^-$ isomeric state~\cite{Byrne2000epja}.

\begin{figure}[!htbp]
\hspace*{-0.07\columnwidth}
\includegraphics[width=1.05\columnwidth]{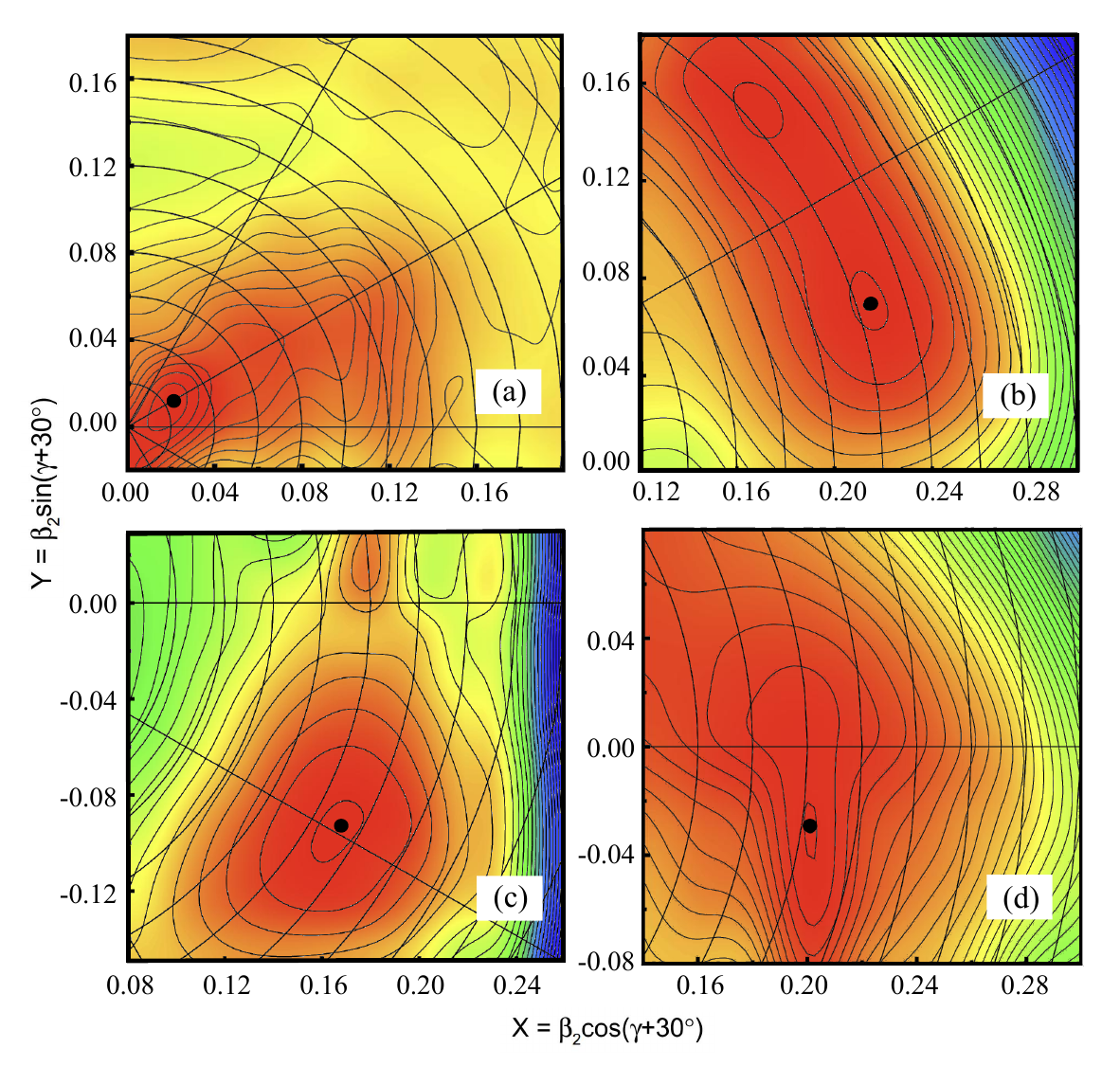}
\caption{Calculated $ { }^{187} \mathrm{Tl} $ potential energy surfaces (PES's) of (a) the near-spherical g.s., (b) the prolate  $K^{\pi}=27/2^+$, $\pi 11/2^{-}[505] \otimes \nu \left\{9/2^{+}[624] \otimes 7/2^{-}[514]\right\}$ state, (c) the oblate $K^{\pi}= 29/2^{+}$, $\pi 13/2^{+}[606] \otimes \nu \left\{9/2^{+}[624] \otimes 7/2^{+}[633]\right\}$ state, (d) the triaxial $K^{\pi}= 25/2^{+}$, $\pi 11/2^{-}[505] \otimes \nu \left\{7/2^{+}[633] \otimes 7/2^{-}[503]\right\}$ state. The energy difference between neighboring contours is 100 keV. The black dots denote the minima of PES's.}
\label{fig2}
\end{figure}

In addition to the high-$\Omega$ proton orbitals mentioned above, other deformation-driving high-$j$ high-$\Omega$ orbitals, including the high-$\Omega$ members of the proton $\pi h_{9/2}$ shell, the high-$\Omega$ members of neutron $\nu h_{9/2}$, $\nu i_{13/2}$, and $\nu f
_{7/2}$ shells, would appear close to the proton and neutron Fermi surfaces at both oblate and prolate sides, respectively. The couplings of these high-$\Omega$ orbitals would form energetically low-lying high-$K$ 3-qp configurations that are polarized to different shapes. We summarize the calculated deformations and excitation energies of possible high-$K$ 3-qp configurations in Table~\ref{table2}. Coexisting different types of intrinsic shape are obtained for a variety of high-$K$ 3-qp configurations from the configuration-constrained PES calculations. Fig.~\ref{fig2} depicts the typical PES's corresponding to the 3-qp configurations with spherical, $\gamma$-soft prolate, oblate, and axially asymmetric shapes, respectively. 

Among them, the lowest prolate high-$K$ 3-qp state given by the configuration-constrained PES is the $K^{\pi}=27/2^+$, $\pi 11/2^{-}[505] \otimes \nu \left\{9/2^{+}[624] \otimes 7/2^{-}[514]\right\}$ configuration. The calculated PES of this configuration is shown in the panel (b) of the Fig.~\ref{fig2}. As we discussed in Sect.~\ref{systematics}, this configuration is most likely to be assigned to the observed isomeric state with excitation energy $E_x = 2584.6$ keV and lifetime $T_{1/2} = 0.69$ $\mu$s. Experimentally, it is found that the 2584.6-keV isomer decays to the $I^{\pi}= 25^+$ member of the rotational band, which is assigned to the low-$\Omega$ $\pi1/2^+[660]$ configuration~\cite{Byrne2000epja,guo2024spectroscopy}. To understand this transition, we also compute the deformation and excitation energy of the low-$\Omega$ $\pi1/2^+[660]$ state (see them in Table~\ref{fig2}). The calculated energy of the $\pi1/2^+[660]$ state is 1216 keV, which is compatible with the estimation of the bandhead energy of the observed rotational band. Note that both the $\pi1/2^+[660]$ and $\pi 11/2^{-}[505] \otimes \nu \left\{9/2^{+}[624] \otimes 7/2^{-}[514]\right\}$ configurations have very soft axially asymmetric deformations of $\gamma\sim 15^{\circ}$. The $\gamma$-soft shape breaks down the $K$-conservation and allows the decay from the $K^{\pi}=27/2^+$ state to the $I^{\pi}= 25^+$ state built on the $\pi1/2+[660]$ configuration. The CCPES calculations predict another $K^{\pi}=27/2^+$ configuration that consists of $\pi 11/2^{-}[505] \otimes \nu \left\{9/2^{-}[505] \otimes 9/2^{+}[633]\right\}$ with a smaller $\beta_2\approx0.18$ and a larger $|\gamma|\approx 30^{\circ}$. However, the large axially asymmetric deformation violates the $K$-conservation, and hence may prohibit the formation of the $K$ isomeric state.   

Another interesting 3-qp state that we predict is the $K^{\pi}=29/2^+$, $\pi 13/2^{+}[606] \otimes \nu \left\{9/2^{+}[624] \otimes 7/2^{+}[633]\right\}$ configuration. As seen in the panel (c) of Fig.~\ref{fig2}, its calculated PES shows a minimum that appears at oblate deformation of $\beta_2\approx0.19$ and $|\gamma|\approx60^{\circ}$. The combination of high $K$ value, axially symmetric shape, and its calculated low energy of $E_{x}=1839$ keV, supports the existence of a long-lived $K$ isomer. Therefore, we suggest that this $K^{\pi}=29/2^+$ configuration could be assigned to the observed isomeric state with a lifetime $T_{1/2}=1.1$ $\mu$s, although the position and spin-parity of this isomer cannot be firmly determined because the $\gamma$ rays linking this isomer to low-lying states are still missing~\cite{Byrne2000epja,guo2024spectroscopy}. More recently, it has been found that this isomer decays to the low-lying $13/2^+$ isomer that lies at 1061 keV~\cite{guo2024spectroscopy}. This implies that the $T_{1/2}=1.1$ $\mu$s isomeric state may be oblate deformed and be composed of the same $\pi13/2^+[606]$ configuration, which is compatible with the predicted $K^{\pi}=29/2^+$, $\pi 13/2^{+}[606] \otimes \nu \left\{9/2^{+}[624] \otimes 7/2^{+}[633]\right\}$ state. We thus propose that two long-lived 3-qp high-$K$ isomeric states with different shapes coexist with intermediate excitation energies in $^{187}$Tl. Further measurement of observables, such as gyromagnetic ratios or electromagnetic transition properties of rotational bands built on these two long-lived states, would help us to unambiguously pin down the shapes and intrinsic structures of these observed isomers.

Other low-lying 3-qp high-$K$ states are predicted by the configuration-constrained PES calculations. Among them, the $K^{\pi}= 25/2^-$, $\pi 9/2^{-}[505] \otimes \nu \left\{9/2^{+}[624] \otimes 7/2^{+}[633]\right\}$ configuration is of particular interest because of its very low energy and high spin value. Since it is calculated to be even energetically lower than the observed lowest $I^{\pi}=13/2^+$ state, the calculated $K^{\pi}= 25/2^-$ state can be expected to form a ``spin trap''~\cite{walker1999energy}. As Dracoulis \textit{et al}.~\cite{dracoulisPLB2012Ir} have
pointed out, a very low energy could result in
long-lived states that would preferentially $\beta$ decay and thus be
missed. It therefore provides a challenge for both experimental and theoretical studies and, in effect, a test of the reliability of the models.

\section{Summary}
\label{sec:summary}

We present a systematic theoretical study of shape polarization and coexistence in high-$K$ 3-qp states of odd-mass $N=106$ isotones ($^{175}$Tm, $^{177}$Lu, $^{179}$Ta, $^{181}$Re, $^{183}$Ir, $^{185}$Au, $^{187}$Tl) using the configuration-constrained PES method.

The investigation focuses on 3-qp states formed by coupling a single proton to the systematic two-quasineutron $K^{\pi}=8^{-}$ isomeric configuration known in the even-even $N=106$ cores. The calculations demonstrate excellent agreement with the experimental data for the well-established isomers in lighter isotones ($Z=69-75$), validating the theoretical approach. As the proton number increases towards the $Z=82$ shell closure, the ground states become progressively softer and less deformed, while the high-$K$ 3-qp states exhibit significant shape polarization, maintaining well-defined prolate deformations. This leads to a substantial shape difference between the isomers and the ground states in nuclei like $^{185}$Au and $^{187}$Tl.

Furthermore, we analyze in detail the intrinsic shapes of 3-qp states in $^{187}$Tl. The configuration-constrained PES calculations identify a multitude of low-lying high-$K$ configurations with distinctly different shapes (including \textit{prolate}, \textit{oblate}, and \textit{triaxial}) coexisting within a narrow energy range. Two specific long-lived isomers observed in $^{187}$Tl are assigned to configurations with different shapes: the $T_{1/2}=0.69 \mu$s isomer is associated with a prolate-deformed $K^{\pi}=27/2^+$ state, while the $T_{1/2}=1.1 \mu$s isomer is proposed to be an oblate-deformed $K^{\pi}=29/2^+$ state characterized by a high $K$ value, an axial shape, and a low excitation energy of 1839 keV, which favors a long lifetime. The study also predicts a very low-lying $K^{\pi}=25/2^{-}$ 3-qp state in $^{187}$Tl, which could act as a ``spin trap'' and presents a challenge for future experimental detection.

In summary, this work presents a systematic description of high-$K$ isomers in the $N=106$ isotonic chain. The calculated excitation energies and deformations are in reasonable agreement with available experimental data, highlighting the crucial role of unpaired nucleons in driving shape polarization and revealing the possible coexistence of distinct shapes in neutron-deficient odd-mass nuclei near the $Z=82$ shell closure. We hope that the present results can provide useful guidance for future spectroscopic investigations in this mass region.

\end{document}